**Controlling Enhancement of Transmitted Goos-Hänchen Shifts: From Symmetric to Unidirectional**

*Zhuolin Wu,[1,2] Weiming Zhen,[1,2] Zhi-Cheng Ren,[1,2] Xi-Lin Wang,[1,2] Hui-Tian Wang [1,2,4] and Jianping Ding,[1,2,3] ***

[1]National Laboratory of Solid State Microstructures and School of Physics, Nanjing University, Nanjing 210093, China

[2]Collaborative Innovation Center of Advanced Microstructures, Nanjing 210093, China

[3]Collaborative Innovation Center of Solid-State Lighting and Energy-Saving Electronics, Nanjing 210093, China.

[4]htwang@nju.edu.cn;

*Corresponding author: jpding@nju.edu.cn

Keywords: Goos–Hänchen shifts, polarization singularities, unidirectional guided resonances, bound states in the continuum

Since the discovery of the Goos-Hänchen (GH) shift in the 1940s, its deep connections to Fourier transforms and causality have led to widespread interest and applications in optics, acoustics, and quantum mechanics. Control of the shift involves both its magnitude and direction. Although resonance-enhanced GH shift under reflection has significantly expanded and facilitated its observation and application, implementations in transmission scenarios remain scarce. More importantly, discussions on the direction of the GH shift are rare, and the associated degree of freedom for controlling directional asymmetry has not been fully explored. To address these issues, we discuss a control framework for enhancing transmitted GH shifts from symmetric to asymmetric. A design with complete degrees of freedom from symmetric shift enhancement to unidirectional shift enhancement is demonstrated in transmission scenarios. The control dimension associated with directionality significantly enhances the flexibility of beam shift control, with broad application prospects in scenarios such as high-sensitivity sensing, precision measurement, optical isolators, and asymmetric optical switches.

**1. Introduction**

The Goos-Hänchen shift, defined as the in-plane spatial shift of the beam's centroid with respect to the plane of incidence at an interface, is an intriguing manifestation of finite-width light beam.

Over the past few decades, extensive research has gradually clarified the physical picture of the GH shift[1–4], which has found broad applications in optical differentiation and imaging[5–7], optical switching[8], as well as sensing and precision measurement[9–12]. The direction and magnitude of the lateral movement of the beam's centroid in space are indispensable for describing and controlling the interaction of light with interfaces. Earlier studies primarily focused on discussions of the displacement magnitude. Typically, the displacement at an interface is on the order of the wavelength, making direct observation extremely challenging[13,14]. While this characteristic has been sought after in fields such as optical computing and edge detection in imaging, it does not fulfill the requirements for applications like sensing. Therefore, enhancing the beam shift has become an important area of research. Among the various enhancement strategies[15–17], resonance-based enhancement, triggered by subwavelength artificial microstructures, stands out as an exceptionally effective approach[9–11,18–29], with the phase discontinuity directly addressing the core of the displacement phenomenon.

With advancements in light-trapping modes, the focus on improving quality factors ($Q$-factors) has mechanistically ensured the sustainability of further enhancement of GH shift. Recently, the complete localization of light in bound states in the continuum (BICs) has significantly strengthened light-matter interactions due to their theoretically infinite $Q$-factors and the topological robustness of momentum space[30–37]. The introduction of this degree of freedom has universally propelled the development of various optical responses[38–46], and the GH shift is no exception, expanding the tunable boundaries of enhancement. However, most of the aforementioned strategies are limited to reflective scenarios. The presence of valleys in transmission scenarios results in low efficiency, restricting the further exploitation of enhanced displacements. Although a targeted solution based on coupled bilayer gratings has been proposed to achieve efficient displacement enhancement in transmission scenarios[22], its dependence on far-field coupling imposes strict implementation requirements, thereby constraining applicability and further improvement. Moreover, and more importantly, it is perhaps because the GH shift is intuitively linked to the plane of incidence that discussions on the direction of displacement are extremely rare. This suggests that the degree of freedom corresponding to directional control in displacement regulation is missing, which limits the versatility of GH shift in more flexible optical field manipulation scenarios.

Our research demonstrates that, under *P*-symmetry protection, the control of radiation for modes propagating in the forward and backward directions can be transformed into the manipulation of radiation in the upward and downward directions for the same mode. As a result, mode radiation control, driven by the *z*-direction mirror symmetry breaking, can further induce an asymmetric GH shift. Based on this, we discuss a control framework for the shift enhancement from symmetric to asymmetric in dislocated layered photonic crystals (DLPC), aiming to fill the research gap related to the missing degrees of freedom in directional coupling control. In this design, non-half-period dislocations in the structure promote the asymmetric enhancement of the shift. Specifically, under full transmission, the enhancement of unidirectional GH shift, as an extreme case of asymmetry, can be supported by unidirectional guided resonance (UGR), which corresponds to the single-side V-point from the perspective of far-field radiation polarization topology[47–53]. Subsequently, we introduce the interlayer spacing degree of freedom to provide an additional dimension for phase control, and the construction of a two-dimensional parameter space leads to a richer evolution of the shift response. Consequently, interference modulation can also be achieved in structures without dislocations or with half-period dislocations, resulting in high-efficiency symmetric shift enhancement. On the other hand, we propose a dislocation-cascade spatial coupling mode theory (DC-SCMT) to accommodate the asymmetric scattering description when the structural mirror symmetry is completely broken. From symmetric to unidirectional shift enhancement, DC-SCMT effectively captures the influence of structural geometric degrees of freedom on scattering behaviors. Furthermore, by examining the phase difference in radiation coupling derived from DC-SCMT, we are able to quantitatively distinguish and characterize, from the scattering perspective, the general interference coupling radiation modes as well as special dark modes, such as BIC and UGR. Our results unlock additional degrees of freedom for the manipulation of interface beams, providing a flexible and developmental approach for transmission shift enhancement. This framework, in turn, broadens the potential applications of the GH shift in practical scenarios such as high-sensitivity sensing, precision measurement, optical isolators, and asymmetric optical switches.

## 2. Results

In contrast to the symmetric beam shift supported by the traditional unperturbed grating in **Figure 1**(a), the introduction of a slight dislocation in the DLPC leads to an asymmetrically enhanced beam shift on either side of the normal. Remarkably, this enables a significant enhancement of the GH shift for single-sided incidence, as illustrated in Figure 1(b). This addresses two issues: the inability to control the degree of directional asymmetry of beam shift and the low transmission at the resonance position that limits the observation and application of the displacement. Furthermore, comparing the insets in Figures 1(a) and 1(b), it is evident that the dislocation operation does not significantly affect the band structure, yet the transmission is substantially improved, with the fundamental transverse electric ($TE_1$) mode transitioning from the hollow red mark to the solid red mark. Although the symmetry in the *z*-direction is broken, the identity of the two stacked grating layers still provides inversion symmetry protection for both the forward and backward modes. This ensures that the topologically addressed UGR will exhibit a full transmission response, represented by the solid red point. This supports the realization of near-unity efficiency, giant enhancement of unidirectional beam shift.

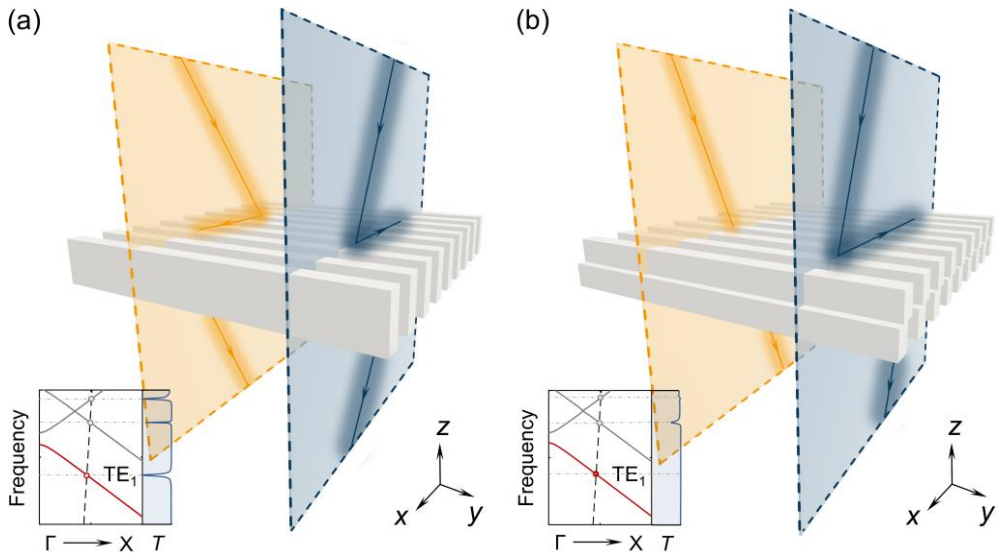

**Figure 1.** Schematic of symmetric and asymmetric GH shift. (a) Symmetry-enhanced shift in a one-dimensional photonic crystal, and (b) unidirectional shift enhancement in a DLPC. The GH shift transitions from symmetric to completely asymmetric with the introduction of dislocations. Blue and orange represent oblique incidence to the right and left of the normal, respectively. The inset shows the corresponding band structure and transmission at a fixed incident angle, with the $TE_1$ band highlighted in red.

## 2.1. Framework for Describing the Control of the GH Shift Enhancement

Under Gaussian illumination, the longitudinal centroid shift of the beam along the propagation direction within the incident plane is expressed as[3]

$$\langle \mathbf{R} \rangle = N^{-1} \langle \tilde{\mathbf{E}}_s | i \frac{\partial}{\partial \mathbf{k}} | \tilde{\mathbf{E}}_s \rangle, \tag{1}$$

where $\tilde{\mathbf{E}}_s$ is the scattered field after the optical system acts on the incident Gaussian mode field $\tilde{\mathbf{E}}_G(k_x, k_y) = w_0^2/(2\pi) \exp\left[-w_0^2\left(k_x^2 + k_y^2\right)/4\right]$ with waist radius $w_0$. $N$ is the the norm defined as $N = \langle \tilde{\mathbf{E}}_s | \tilde{\mathbf{E}}_s \rangle$. This reflects the direct correspondence between the phase gradient experienced by the angular spectrum distribution (i.e., the wavevector distribution in momentum space) and the real-space displacement of the beam's center of mass. When the beam's lateral phase variation is gradual with respect to the scatterer, Equation (1) can be approximated as

$$R_x = -\frac{\partial \varphi}{\partial k_x}, \tag{2}$$

to describe the in-plane displacement $R_x$ under oblique incidence in the $xz$ plane, where $\varphi$ is the phase of the transmission coefficient. In many scenarios, this approximation helps us conveniently describe and analyze the shift behavior of beams.

To describe the transmission coefficients near Γ point in momentum space, we develop a theoretical framework as DC-SCMT to accommodate the broken mirror symmetry, while being compatible with the traditional SCMT originally designed for mirror symmetric systems[25,26,44]. This improvement adapts the theory to account for the tunability of dislocation and spacing in the dual-layer and even multi-layer gratings of the DLPC structure. The propagation characteristics of mode $A^{(j)}(\mathbf{k}_\parallel, \mathbf{r}) = u^{(j)}(\mathbf{k}_\parallel, \mathbf{r}) \exp(i\mathbf{k}_\parallel \cdot \mathbf{r}) \exp(-\alpha_{v_g}^{(j)} \gamma^{(j)} \mathbf{r}_\parallel)$ in the $j$-th layer photonic crystal (where $j \in \mathbb{N}$) can be described by a four port model

$$\begin{aligned} \frac{d\mathbf{A}}{d\mathbf{r}_\parallel} &= \mathbf{H}_0 \mathbf{A} + \boldsymbol{\alpha}_{v_g} \mathbf{K}^T | s_{in} \rangle, \\ | s_{out} \rangle &= \mathbf{C} | s_{in} \rangle + \mathbf{D} \boldsymbol{\alpha}_{v_g} \mathbf{A} = \mathbf{S} | s_{in} \rangle, \end{aligned} \tag{3}$$

where $\mathbf{A} = (A^{(j)+}, A^{(j)-})$ is the mode amplitudes of forward and backward propagation, and $\boldsymbol{\alpha}_{v_g} = \text{diag}\left(\alpha_{v_g}^{(j)+}, \alpha_{v_g}^{(j)-}\right)$ represents the relationship between the mode group velocity and dispersion ($\alpha_{v_g}^{(j)} = 1$ for $v_g > 0$ and $\alpha_{v_g}^{(j)} = -1$ for $v_g < 0$). Here, $\mathbf{r} = (x, y, z)$ denotes the position vector, $\mathbf{r}_\parallel = (x, y)$ its in-plane component parallel to the in-plane wavevector $\mathbf{k}_\parallel = (k_x, k_y)$. Thereupon, the scattering matrix linked to the input $|s_{in}\rangle$ and output $|s_{out}\rangle$ is denoted as

$$\mathbf{S} = \mathbf{C} + \mathbf{D}\mathbf{H}^{-1}\boldsymbol{\sigma}_x \mathbf{D}^T \tag{4}$$

where $\boldsymbol{\sigma}_x$ is the first in the set of Pauli matrices. The direct scattering coefficient is modified by the influence of nearby band structures, incorporating a background resonance term[49]

$$\mathbf{C}_b = \begin{pmatrix} 0 & r & 0 & it \\ r & 0 & it & 0 \\ 0 & it & 0 & r \\ it & 0 & r & 0 \end{pmatrix} = \mathbf{C}_0 - \frac{d_b^2}{i(k_\parallel - k_b) \pm \gamma_b}, \mathbf{C}_0 = \exp(i\varphi) \begin{pmatrix} 0 & r_0 & 0 & it_0 \\ r_0 & 0 & it_0 & 0 \\ 0 & it_0 & 0 & r_0 \\ it_0 & 0 & r_0 & 0 \end{pmatrix} \quad (5)$$

where $\mathbf{C}_0$ describes the Fabry-Pérot resonance[54], and $d_b$, $k_b$, and $d_b$ represent the radiative coupling coefficient, the eigen wave vector, and the loss associated with the background collective resonance, respectively. The Hamiltonian $\mathbf{H}_0 = i|\mathbf{k}_\parallel|\sigma_z - \mathbf{H}$ governs DLPC with the third Pauli matrix $\sigma_z$. $\mathbf{K}$ and $\mathbf{D}$ indicative are coefficients of the coupling during scattering, which are constrained by $\mathbf{D}^\dagger\mathbf{D} = 2\gamma\mathbf{I}$, $\mathbf{K} = -\mathbf{D}\sigma_x\mathbf{\alpha}_{v_g}$, $\mathbf{CD}^* = -\mathbf{D}\sigma_x$. For details and derivations, refer to the subsequent discussions and **Supporting Information**. Simultaneously, the transfer matrix associated with the transmission and reflection coefficients, can be obtained as[55]

$$\mathbf{T}^{(j)} = \begin{pmatrix} \left(\frac{1}{t^{(j)}}\right)^* & \frac{r^{(j)}}{t^{(j)}} \\ \left(\frac{r^{(j)}}{t^{(j)}}\right)^* & \frac{1}{t^{(j)}} \end{pmatrix}. \quad (6)$$

When $j > 1$, the phase $\theta$ attributed to the optical path difference between adjacent layers is described by the propagation matrix $\mathbf{P}^{(j-1)} = \mathrm{diag}[\exp(i\theta^{(j-1)}), \exp(-i\theta^{(j-1)})]$[55], so that the overall transmission and reflection coefficients of the system can be extracted from the complete transfer matrix

$$\mathbf{M} = \prod_{j=1}^{N} \mathbf{T}^{(j)} \mathbf{P}^{(j-1)} = \begin{pmatrix} M_{11} & M_{12} \\ M_{21} & M_{22} \end{pmatrix},$$
$$t_{\mathrm{all}} = \frac{1}{M_{22}}, r_{\mathrm{all}} = \frac{M_{12}}{M_{22}}. \quad (7)$$

where $\mathbf{P}^0$ is defined as identity matrix $\mathbf{I}_2$. With this formulation, the GH shift can be effectively analyzed. To intuitively describe the directional selectivity of radiation from the coupled modes in an $N$-layer DLPC from a scattering perspective, we define the phase difference within the scattering channels that compete for radiation directionality, with the layer numbering starting from 1 at the bottom and increasing upwards. This is primarily estimated through contributions from three parts:

$$\psi = (\psi_0 + \psi_w + \psi_h) \bmod 2\pi, \quad (8)$$

where $\psi_0$ is the inherent phase difference of the radiation coupling coefficients between different layers in the scattering channel. For $j > 1$, $\psi_0$ is given by $\arg(d^{(1)}) - \sum_{j=2}^{N} \arg(d^{(j)})$, where $d^{(j)}$ is the complex coupling coefficient of the $j$-th layer. Notably, for $j = 1$, the radiation reduces to that from a single layer, with the phase of the port coupling coefficient directly corresponding to $\psi_0$. $\psi_w = \mathbf{k} \cdot \mathbf{\delta}_w$ and $\psi_h = \mathbf{k} \cdot \mathbf{\delta}_h$ represent the phase induced by the transverse structural dislocation ($\delta_w$) and the phase introduced by the interlayer spacing ($\delta_h$) in the longitudinal direction, respectively. In the transmission process, we choose the positive direction of $\delta_w$ as $+x$, while $-z$ is set as the positive of $\delta_h$. The wave number $k = n_{eff}k_0$ is related through the effective refractive index $n_{eff}$ to the vacuum wave number $k_0$. To quantitatively evaluate the asymmetry in the displacement response, we define an asymmetry factor (AF) as

$$\text{AF} = \frac{\left||R_x^+| - |R_x^-|\right|}{\left||R_x^+| + |R_x^-|\right|} \tag{9}$$

Subsequent discussions will address the specific settings and applications under specific scenarios, providing support for designing asymmetric beam shifts.

## 2.2. Dislocation-Induced Asymmetric to Unidirectional GH Shift Enhancement

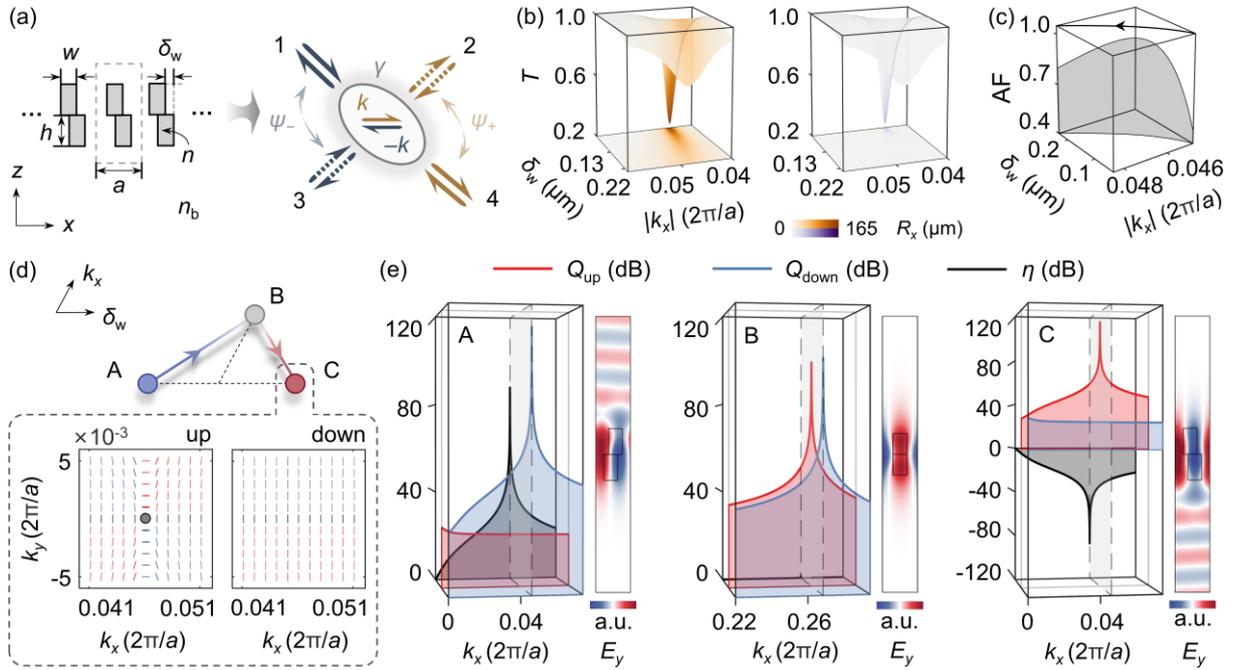

**Figure 2.** Scattering and mode analysis during dislocation evolution. (a) Schematic of the stacked DLPC structure (left) and its abstract dual-resonance four-port model (right). $\psi_+$ and $\psi_-$ denote the phase differences between the upper and lower side radiations of forward and backward modes induced by dislocations. (b) Transmission and displacement distributions in the parameter space defined by the dislocation and incident wave vector components, with left and right panels representing rightward and leftward oblique incidence, marked in orange and purple, respectively. (c) Mode evolution in parameter space with increasing the dislocation and the corresponding AF of the induced GH shift. (d) Mode evolution from upward-radiating UGR (Mode A) to off-Γ BIC (Mode B) and then to downward-radiating UGR (Mode C), with the inset showing the far-field polarization topology of momentum space for Mode C (red: LCP,

blue: RCP, and black: linear polarization). (e) *Q*-factors and radiation asymmetry for Modes A, B, and C, with the inset displaying the field distribution at each gray cross-section.

Symmetry breaking in the GH shift enhancement, caused by interlayer dislocations of the photonic crystal slabs, manifests when the beam is incident on either side of the normal. To systematically investigate the influence of dislocations on the asymmetric displacement, we designed a stacked DLPC (**Figure 2**(a)), which is embedded in a SiO$_2$ environment ($n_b$ = 1.5). The structure comprises a grating with a width of $w$ = 0.4$a$ and a height of $h$ = 0.75$a$, periodically arranged with a period $a$. The refractive index of the grating material is $n$ = 2 (Si$_3$N$_4$). When the parameter $\delta_w \neq 0$, the $z$-reflection symmetry of the structure is broken, while inversion symmetry (*P* symmetry) remains preserved due to the consistency of the two layers. In our theory, when the interlayer distance is very small, a reliable approach is to treat the structure as a single-mode grating. This avoids the violation of the small perturbation approximation typically required in conventional coupled-mode theory[56]. Also, as contrasted with time-domain discussions, forward- and backward-propagating modes cannot be merged into a single mode description, even though they share the same characteristic frequency and loss rate[57]. Due to their opposite group velocities, they must be represented in a matrix form as

$$\mathbf{H}_0 = i|\mathbf{k}_0|\boldsymbol{\sigma}_z - \boldsymbol{\alpha}_{v_g}\gamma \tag{10}$$

where $|\mathbf{k}_0|$ is the mode wavenumber, and $\gamma$ represents the radiation loss. It is worth noting that the radiation loss here accounts for the effects of both upward and downward radiation. We do not consider it convenient to separate these contributions. Instead, we incorporate the asymmetry in radiation caused by dislocation into the radiation coupling coefficient matrix

$$\mathbf{D} = \begin{pmatrix} 0 & d_1 \exp(i\psi_-/2) \\ d_2 \exp(-i\psi_+/2) & 0 \\ 0 & d_3 \exp(-i\psi_-/2) \\ d_4 \exp(i\psi_+/2) & 0 \end{pmatrix} \tag{11}$$

where $d_p$ ($p$ = 1-4) represents the four-port radiation coupling coefficient, with $d_1 = d_4 = d'$, $d_2 = d_3 = d''$ as well as $|d'|^2 + |d''|^2 = 2\gamma$. As a core concept, the introduced port phase $\psi_+ = \psi_- = \psi_w$ in Equation (8) acts as a bridge linking the geometric dislocation of the structure to the asymmetric radiation in DC-SCMT. Although $\theta$ does not contribute to the longitudinal phase difference, the optical path difference within the layer must be considered. Following the theoretical framework outlined in Section 2.1, we can effectively describe the GH shift.

As the incident light is obliquely incident from either side of the normal, the transmission coefficients and the corresponding displacement distributions in the parameter space, determined by the wavevector components and dislocations, are illustrated in Figure 2(b). According to Equation (2), asymmetric GH shifts can be estimated, where rightward oblique incidence experiences enhancement (left panel), while the other does hardly any (right panel). The corresponding transmission coefficients remain identical. Additionally, the transmission dip disappears during its evolution, indicating the potential for significant displacement enhancement under all-pass conditions. Based on Equation (9), we calculated the evolution of the AF of the beam shift along the trajectory of the mode in parameter space, as shown in Figure 2(C). The AF exhibits a trend of initially increasing and then decreasing. In the high-transmission region, the AF approaches 1, quantitatively demonstrating the characteristic of unidirecitonal enhancement. The GH shift at theoretical perfect transmission is characterized by near-unity efficiency. Conversely, beam shifts with lower AF will encounter difficulties in observation due to reduced efficiency.

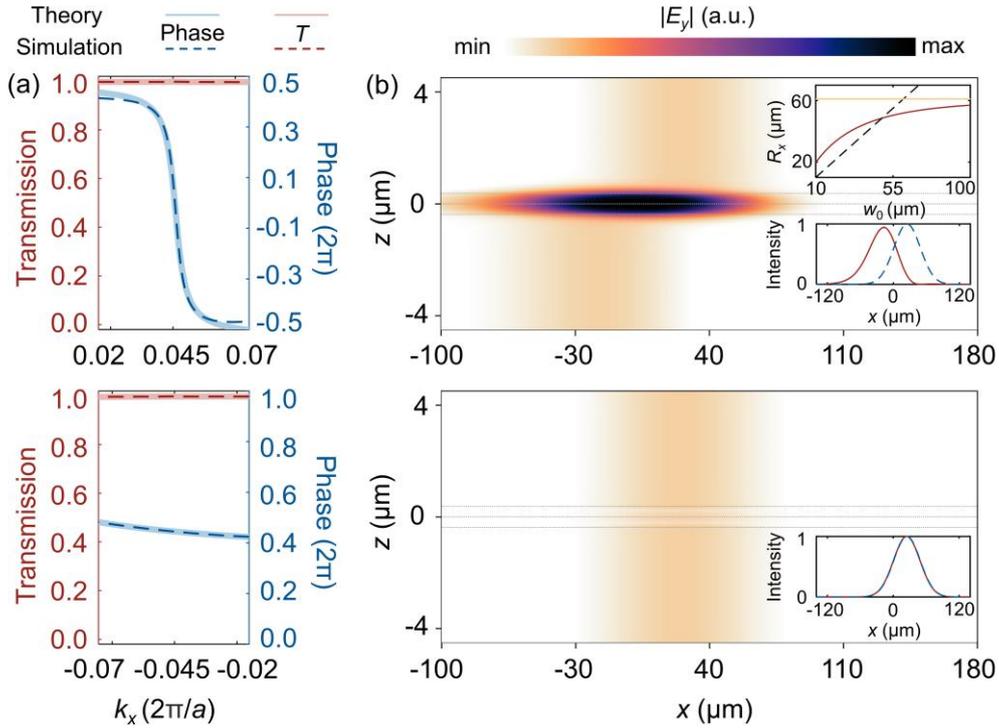

**Figure 3.** Dislocation-induced unidirectional GH shift enhancement. (a) Angular spectrum of transmission and phase for Mode C at normalized frequency 0.5774, with solid and dashed lines for theoretical and simulation results, respectively. Upper panel: $k_x > 0$; lower panel: $k_x < 0$. (b) Field distribution and far-field intensity slices for oblique incidence with wave vector components $k_x a/2\pi = \pm 4.6120 \times 10^{-2}$, shown on the upper and lower panels, respectively. The red solid and blue dashed lines show the responses from the incident and transmitted ports, respectively. The dashed line delineates the boundaries of each layer in the DLPC. The inset illustrates the shift variation with the Gaussian beam waist radius, marked in red, with yellow indicating the approximated shift, and the black dashed line representing the contour where the waist radius equals the shift.

From the perspective of mode analysis, we explored the changes in eigenmodes corresponding to scattering responses during the evolution of the dislocation. In regions with a high AF, a UGR has been identified. As shown in the inset of Figure 2(d), the mode exhibits a one-sided V-point topology in momentum space, indicating radiation only to one side. The investigated forward-propagating along $x$ (+$k$), labeled as C, suppresses upward radiation, leading to a significant transmission shift enhancement in the rightward oblique incidence due to the presence of downward radiation. Conversely, by $P$ symmetry, the leftward oblique incidence does not exhibit displacement enhancement, as the $-k$ mode suppresses downward radiation. Unfortunately, due to interference from modes corresponding to other bands during the scattering process, the AF of GH shift rarely achieves an ideal value of 1, and the parameters corresponding to UGR excitation show slight deviations. These non-ideal effects become more pronounced as the bands approach each other. Figure 2(d) illustrates the evolution of the +$k$ mode during the dislocation evolution process. As the dislocation $\delta_w$ evolves from −0.1357$a$ to 0.1357$a$, an upward-radiating UGR (Mode A) initially moves away from the Γ point and, in the absence of dislocation, forms an off-Γ BIC (Mode B) with completely suppressed radiation. It then evolves into a downward-radiating UGR (Mode C) that is consistent with the wave vector of Mode A, but with the radiation and non-radiation sides swapped. Quantitatively, the $Q$-factor for the upward- and downward-radiating modes[51], along with the asymmetry ratio $\eta = Q_{up}/Q_{down}$, reflect the UGR corresponding to ($\delta_w/a$, $k_x a/2\pi$) = (±0.1357, 4.6120×10$^{-2}$), in contrast to the BIC corresponding to (0, 2.6848×10$^{-1}$).

We now select Mode C from Figure 2(d), which is the downward-radiating UGR, for further verification. **Figure 3**(a) illustrates the scattering characteristics of the mode, demonstrating excellent agreement between the theoretical and simulation results based on the DC-SCMT. Specifically, for the transmission response of Mode C under excitation at Port 1, we observe a phase difference of $\psi = 6.2801$ for the downward radiation channel during forward propagation, which reflects the constructive interference feature. Conversely, through transmission on the opposite side, the phase difference of 3.0653 is obtained, corresponding to the upward radiation channel of the forward mode under *P*-symmetry protection, which satisfies the condition for destructive interference. It is evident that the energy redistribution between the two scattering channels of the mode aligns well with the radiation characteristics of the UGR. For a Gaussian light source with oblique incidence, the transverse phase variation becomes significant when the beam waist is small, making the approximation in Equation (2) inaccurate. As shown in the inset of Figure 3(b), the GH shift increases with beam waist radius (red solid line), while asymptotically approaching 60.96 μm as predicted by the phase gradient approximation (yellow solid line). From an observational perspective, the relationship between the GH shift and the beam waist size is non-negligible. The contour lines (black dashed line) divide the space into two regions, with the upper and lower sides corresponding to displacements greater than or less than the beam waist radius, respectively. Therefore, to achieve sufficiently large displacement while remaining convenient for observation, we choose a beam waist radius of $w_0 = 48$ μm, near the intersection of the red solid line and black dashed line. The response field distribution cross-section and intensity slices for this Gaussian light incidence are shown in Figure 3(b). A substantial enhancement in the unidirectional transmitted GH shift is observed, reaching 48.3478 μm on the radiative side, while the non-radiative side only exhibits a conventional GH shift on the order of the wavelength, without enhancement of mode. The AF is calculated to be as high as 0.9600.

**2.3. Extended Design Flexibility Introduced by Interlayer Spacing**

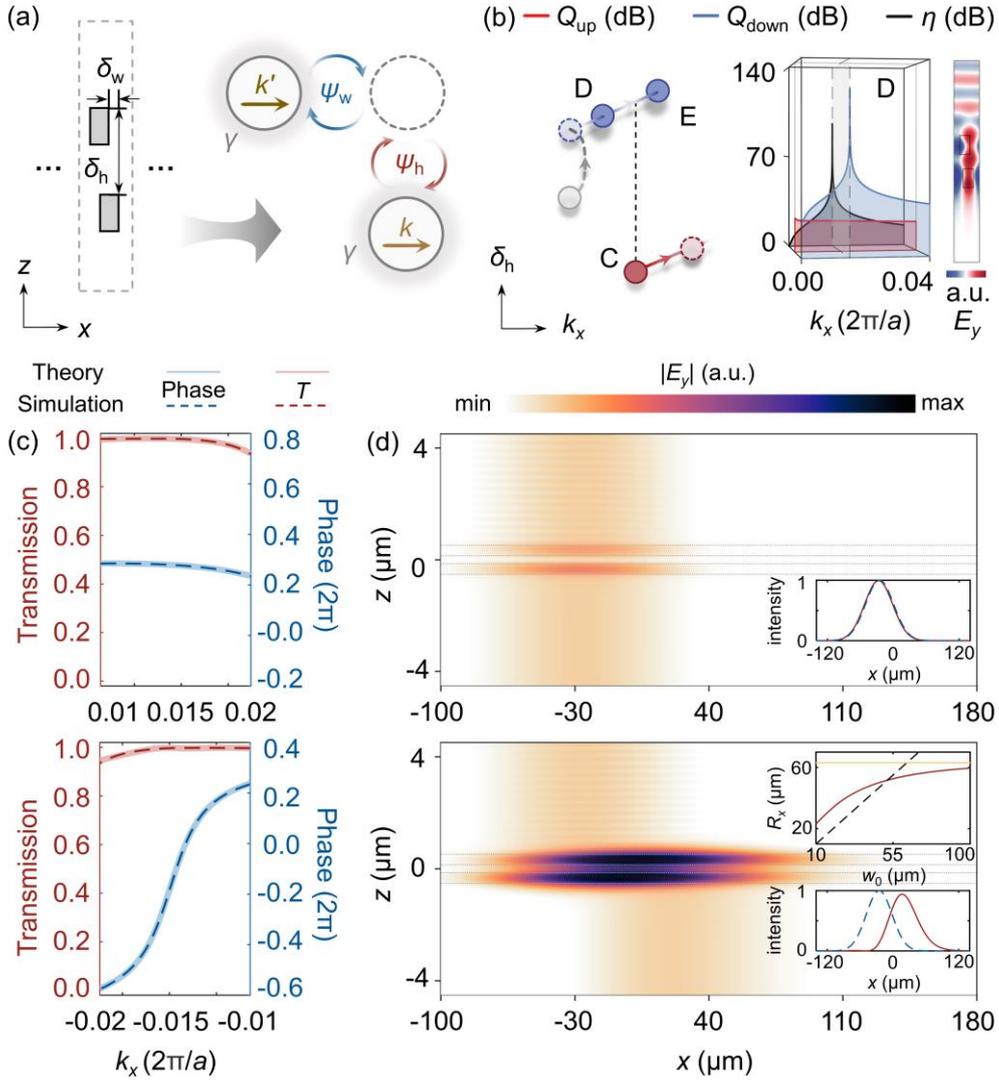

**Figure 4.** Eigenmode and scattering analysis during interlayer spacing evolution. (a) Schematic of the DLPC structure with interlayer spacing (left) and its abstract dual-resonance far-field coupling model (right). $\psi_w$ and $\psi_h$ represent phase differences between modes of layers induced by interlayer dislocations and spacing. (b) Left: Evolution of upward-radiating UGR (Modes D and E) from Mode C as interlayer spacing changes, with dislocation fixed. Right: Q-factor, radiation asymmetry, and field distribution of Mode D. (c) Angular spectrum of transmission and phase for Mode D at normalized frequency 0.6116, with solid and dashed lines for theoretical and simulation results, respectively. Upper panel: $k_x > 0$; lower panel: $k_x < 0$. (d) Field distribution and far-field intensity slices for oblique incidence at $k_x a/2\pi = \pm 1.493 \times 10^{-2}$ for Mode D shown on the upper and lower sides, respectively, with the inset showing the shift variation with Gaussian beam waist radius, using the same color settings as in Figure 3(b).

To further explore the degrees of freedom for controlling GH shift, the interlayer spacing($\delta_h$) emerges as a critical dimension parameter. When $\delta_h$ is relatively large, corresponding to the *j* >

1 case in DC-SCMT as shown in **Figure 4**(a), the DLPC can be described through far-field coupling between multiple identical photonic crystals. In this regime, the mirror symmetry of the radiation ports in each layer remains well-preserved enabling the system to degenerate from a four-port to a two-port symmetric radiation system. This simplification eliminates the need for simultaneous consideration of forward and backward propagation modes. The corresponding Hamiltonian and coefficient matrix are expressed as

$$H_0 = i|k_0| - \alpha_{v_g}\gamma, \; H = i|k_\parallel| - H_0,$$
$$\mathbf{C} = \begin{pmatrix} r & it \\ it & r \end{pmatrix} = \mathbf{C}_0 - \frac{d_b^2}{i(|k_\parallel| - |k_b|) \pm \gamma_b}, \; \mathbf{C}_0 = \exp(i\varphi)\begin{pmatrix} r_0 & it_0 \\ it_0 & r_0 \end{pmatrix} \quad (14)$$
$$\mathbf{D}^{(1)} = \begin{pmatrix} d_u \\ d_d \end{pmatrix}, \; \mathbf{D}^{(2)} = \mathbf{D}^{(1)}\exp(i\psi_0)\exp(i\psi_w), \; \mathbf{K}^{(j)} = -\alpha_{v_g}\mathbf{D}^{(j)}.$$

with $\psi_+ = \psi_- = 0$ and trivial radiation coupling coefficient $|d_u| = |d_d| = \sqrt{\gamma}$. Here, $\psi_0$ represents the phase difference between layers while $\psi_w$ denotes the transverse phase induced by dislocations. From Equation (3), the single-layer photonic crystal scattering matrix can be derived as $\mathbf{S} = \mathbf{C} + \mathbf{D}\mathbf{D}^T/H$. The interlayer optical path is described by $\theta = \psi_h - n_{eff}hk_z$, based on the $\mathbf{P}^{(j-1)}$. The left panel of Figure 4(b) illustrates the mode evolution in parameter space with increasing interlayer spacing under fixed dislocation. Interestingly, at $(\delta_h/a, k_x a/2\pi) = (0, 4.6120\times10^{-2})$, Mode C moves away from Γ-point as the layer spacing increases, while Mode D emerges at (0.3820, 0.0149) with opposing to radiation direction - distinct in origin from Mode C. Near $\delta_h = 0.4309a$, we observe Mode E (a UGR) sharing of Mode D's radiation direction. Band structure analysis (**Supporting Information**) reveals that the scattering response of Mode D experiences significant $TE_2$ band interference, which becomes more pronounced as the layer spacing increases. We therefore focus on Mode D for its realtive isolation from such interactions. The analysis of *Q*-factor and radiation contrast confirms Mode D as an upward-radiating UGR, with field distribution visually validatinh this radiation characteristic ( Figure 4(b) right panel). Under external exciation, the angular spectrum results (Figure 4(c)) demonstrate preserved high transmission for both sides with suppressed radiation phase shift for rightward oblique incidence and evident nearby mode interference. Theoretical and simulation results show satisfactory agreement. For the forward-propagating mode, downward and upward radiation channels exhibit phase difference of 3.6749 and 5.6240, respectively, corresponding to destructive and constructive interference, which indicate the UGR behaviors of Mode D. Figure 4(d) illustrates the illumination effect of a Gaussian beam incident obliquely with a matched resonance wavevector. Similar to the consideration in Section 2.2, we select a beam waist radius of 47 μm (corresponding to the maximum GH shift in the displacement-dominated regions). We observe for the leftward oblique incidence a significant shift enhancement absent in rightward oblique incidence, achieving asymmetric GH shift enhancement. The resulting AF is calculated as |47.2226 μm − 1.4209 μm| / (47.2226 μm + 1.4209 μm) = 0.9416.

### 2.4. High-Efficiency Symmetric Shift Enhancement via Accidental Interference

Recent advances in manipulating GH shift enhancement through coupling between double-layer gratings have revealed significant physical insights[22], showing that precise tuning of interlayer coupling enables remarkable transmission efficiency improvement. However, the specific requirements for half-period dislocation configurations ($\delta_w = \pm 0.5a$) impose stringent constraints on the manipulation of directional asymmetric GH shift. Modal analysis (shown in **Figure 5**(a) and **Supporting Information**) reveals the equivalence between half-period dislocations and non-dislocated configurations under the $z$-reflection symmetry condition, thereby eliminating directional asymmetric degrees of freedom in GH shift modulation. Our comprehensive physical model systematically investigates this unique scenario through parameter-space exploration. When maintaing $\delta_w = 0.5a$ and progressively increasing interlayer spacing $\delta_h$, the band structure evolution along $k_x$ direction exhibits distinctive features (Figure 5(b)): (1) The $TE_1$ and $TE_2$ band initially approach to each other then diverge, with multi-mode interference dominating the coupling region where the two bands are very close; (2) An off-Γ BIC emerges at $\delta_h = 1.35a$ (blue highlighted section), exhibiting the bound state characteristics of BIC modes as shown in Figure 5(c). This BIC manifestation corresponds to complete radiation suppression in scattering response, as shown in Figure 5(d), where only the mode response corresponding to $TE_2$ is observed. The agreement between the transmission and phase behavior, as well as the radiation suppression at a BIC corresponding to the phase difference of $\psi = 3.0212$, both serve as supporting evidence of the validity of our theoretical framework. Notably, the high-transmission regime at at $\delta_h = 1.09a$ (red section in Figure 5(b) corresponds to the high transmission scenario caused by multi-mode interference.

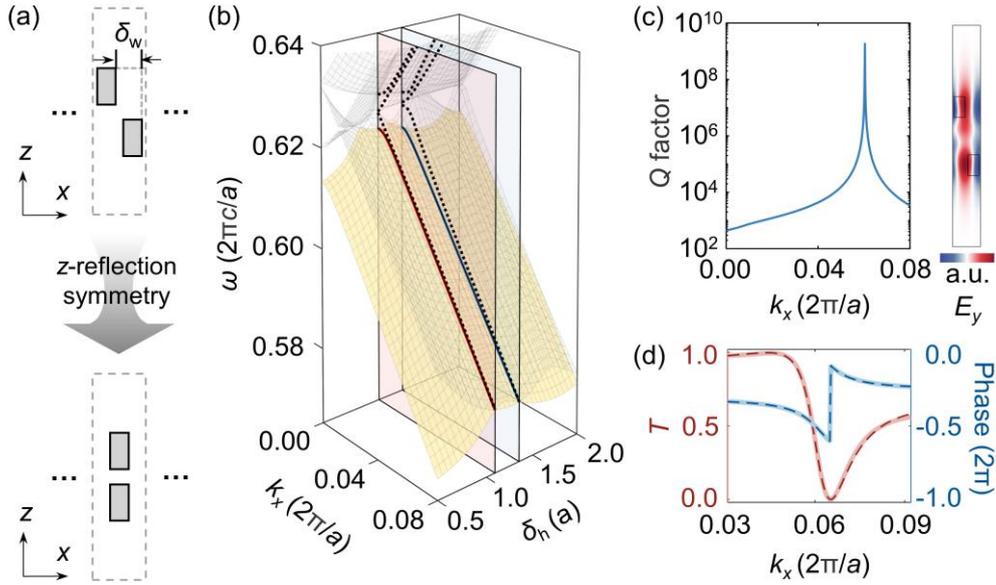

**Figure 5.** Multimode interference under half-period dislocation. (a) Schematic of the DLPC structure with half-period dislocation. (b) Evolution of the band structure with interlayer spacing, with the TE$_1$ band highlighted in yellow. (c) Q-factor distribution for TE$_1$ along the blue line in (b), representing the intersection of the blue section and TE$_1$, with the inset showing the off-Γ BIC field distribution revealed by the infinite Q-factor. (d) Angular spectrum of transmission and phase for the BIC mode, with solid and dashed lines for theoretical and simulation results, respectively.

Specifically, in **Figure 6**, we discuss the scattering response in detail. When the interlayer spacing $\delta_h$ is fixed and the dislocation increases, the transmission for oblique incidence in both directions initially increases then decreases in the parameter space, while the maximum beam shift follows intersecting evolutionary trajectories, reflecting the competition between the two modes. Notably, the theoretical GH shift limit under multi-mode interference remains significantly lower than single-mode scattering effects due to weaker light confinement. Unlike single-mode-induced shifts (single evolutionary trajectory in parameter space), Figure 6(b) plots the displacement evolution supported by multi-mode interference for rightward and leftward oblique incidence in, labeled in yellow and purple, respectively. These two trajectories intersect at $(\delta_w/a, k_x a/2\pi) = (5.0709 \times 10^{-2}, 0.5)$, where the AF on each trajectory approaches zero, indicating the occurrence of high transmission and perfectly symmetric GH shifts at this point. Furthermore, the angular spectrum of the transmission at the intersection is shown in Figure 6(c), where both the transmittance and phase variation align with the predicted characteristics. At this point, the DC-SCMT reveals that both the downward and upward radiation chanels of the forward-propagating mode exhibit constructive interference, with phase differences of 5.9781 and 0.6671, respectively. In contrast to the energy redistribution between scattering channels during single-mode radiation, this results from the competition between modes involved in multi-mode interference. The radiation channels, protected by $z$-reflection symmetry, correspond to the characteristics of a symmetric radiation mode. Additionally, multimode interference can considerably diminish the total radiation at the ports, but energy conservation prevents it from achieving full transmission as in single-port energy exchange, even when the necessary condition of $P$-symmetry in the scattering system is met. Therefore, in contrast with the previously discussed unidirectional GH shift, symmetric shifts can only be achieved with high efficiency. These symmetric shifts depend on different interference processes, but both are accidental singularities in the two-dimensional parameter space. Considering the stricter constraints imposed by the Gaussian light source, to better facilitate experimental observation, we set the beam waist radius to $w_0 = 30$ μm, even though the theoretical displacement limit of 39.0902 μm cannot be fully reached. As expected, both rightward and leftward oblique incidence yield symmetric beam displacement responses, with the AF calculated as |30.7991 μm − 30.7998 μm| / (30.7991 μm + 30.7998 μm) = $1.1364 \times 10^{-5}$, as shown in the field distribution cross-section and intensity slices in Figure 6(d).

## 3. Discussion and Conclusion

Traditional GH shift responses are typically symmetric. By breaking this symmetry, we demonstrate the tunability of GH shift with bidirectional asymmetry as an additional degree of freedom. In light of this intriguing physical insight and the extended theoretical framework, three key remarks are provided here.

First, from symmetric to unidirectional enhancement, we emphasize the full range of tunability range from 0 to 1. Contrary to the conventional intuition of the GH shift, asymmetry is often ubiquitous, while symmetric and unidirectional enhancement appear as singularities sparsely distributed within the parameter space. To fully understand asymmetric beam shift enhancement, we interpret these responses by analyzing scattering from a mode perspective. Moreover, our explanation is also applicable to previous creative implementations based on multimodal interference.

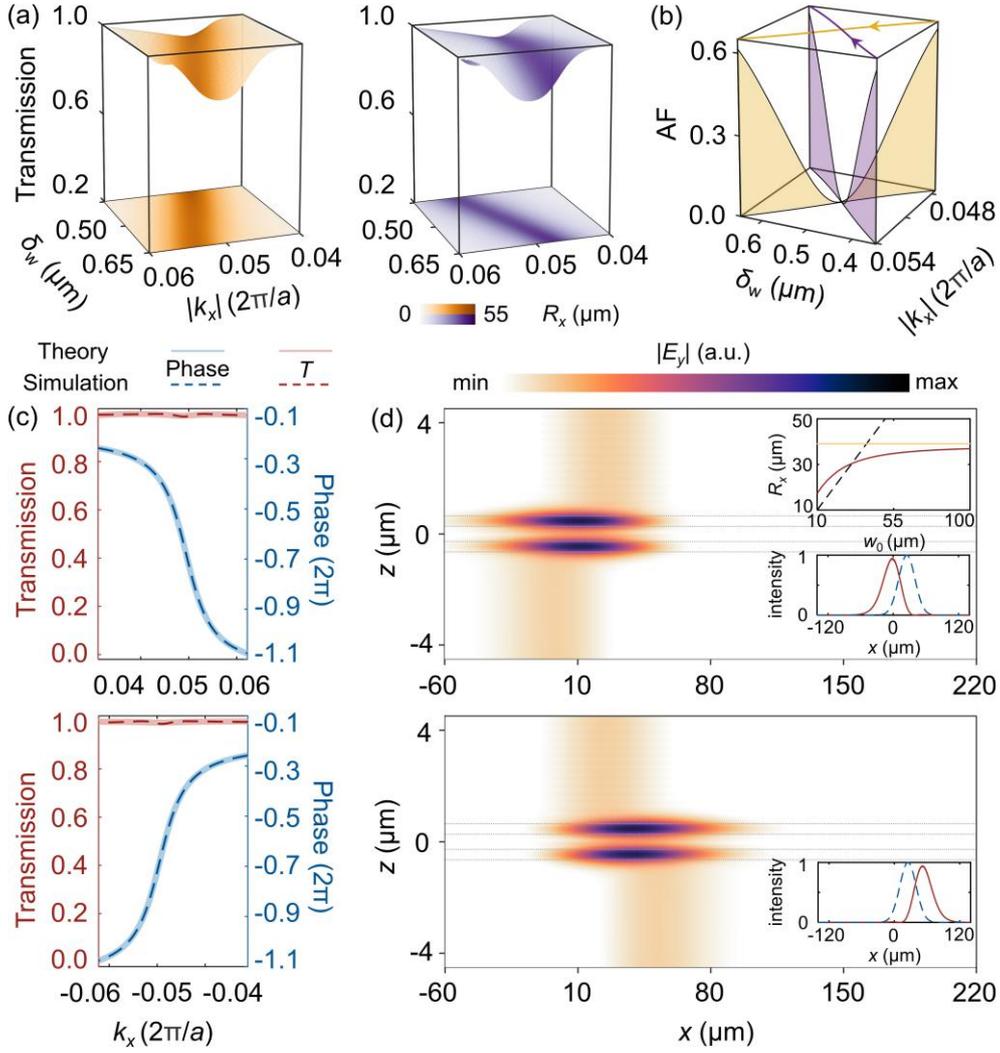

**Figure 6.** Scattering analysis under multimode interference, with interlayer spacing fixed at the red section in Figure 5(b), normalized frequency 0.5945. (a) Transmission and displacement distributions in the parameter space of dislocation and incident wave vector components, with left and right panels representing rightward and leftward oblique incidence, marked in orange and purple, respectively. (b) Mode evolution in parameter space with increasing dislocation and the corresponding AF of the induced GH shift, with incident direction colors matching (a). (c) Angular spectrum of transmission and phase at fixed incident frequency, with solid and dashed lines for theoretical and simulation results. Upper panel: $k_x > 0$; lower panel: $k_x < 0$. (d) Field distribution and far-field intensity slices for oblique incidence with wave vector components $k_x a/2\pi = \pm 0.0490$, with the inset showing the shift variation with $w_0$, using the same color settings as Figure 3(b).

Second, as a formal theory applied to DLPC, the DC-SCMT overcomes two major limitations:

a. Traditional SCMT relies on the stringent assumption of in-plane mirror symmetry within the structure[26,43]. We have relaxed this requirement, extending its applicability to optical systems where in-plane mirror symmetry is broken, while inversion symmetry is preserved.
b. Previous analyses of multilayer systems primarily depended on far-field coupling, where the weak coupling approximation required sufficient separation between layers[52,56]. By treating closely spaced layers as a single effective layer and incorporating the concept of transfer matrices, we extend the theory to a unified formalism applicable to systems with an arbitrary number of layers.

Third, in the pursuit of GH shift enhancement, the implementation in transmission scenarios encounters inherent challenges compared to the relatively well-studied reflection configurations. Our approach introduces two geometric degrees of freedom, enabling flexible design for transmission enhancement and control over directional symmetry, thereby broadening the potential applications across diverse scenarios. As demonstrated in the **Supporting Information**, we show the feasibility of achieving high-efficiency and high-sensitivity sensors based on unidirectional transmission shift enhancement. Moreover, it is worth noting that for asymmetric enhancement, layer spacing is not a strict requirement, meaning fewer degrees of freedom are needed, leading to simpler and more adaptable designs. On the other hand, under the condition of constant total scattered radiation, energy conservation ensures that unidirectional radiation gains a higher proportion through interference redistribution. This results in a significantly larger shift response compared to those induced by general asymmetric or symmetric enhancement. Furthermore, the growing availability of rich high-$Q$ modes[36,58,59] offers exceptional scalability for our GH shift enhancement framework, paving the way for broader applications and design flexibility.

In summary, we propose a design for GH shift enhancement from symmetric to asymmetric, specifically discussing the evolution of beam displacement under the control of two designable degrees of freedom: dislocation and interlayer spacing. The near-unity-efficient unidirectional GH shift enhancement under theoretical full transmission and high-efficiency symmetric shift enhancement based on DLPC are typical responses in the parameter space. Furthermore, we have developed DC-SCMT for structures with broken mirror symmetry, providing a unified formal description of the scattering behavior in the parameter space. This theory establishes a bridge between the geometric perturbations in structural design and the response of radiation. Our approach introduces an additional design dimension for GH shift enhancement, significantly increasing the design flexibility for achieving giant transmitted GH shifts. This makes the observation and utilization of displacement more accessible, with potential applications in high-sensitivity sensing and precision measurement, optical switches, as well as optical isolators.

**4. Methods**

The eigenmode analysis and far-field polarization calculation of the DLPC are performed based using the finite-element method. In the two-dimensional modeling within the $x$-$z$ plane, periodic boundary conditions are applied in the $x$-direction, while perfectly matched layers (PMLs) are used in the $z$-direction for absorbing outgoing waves. The polarization vector of the resonance in the far field is defined through the Fourier transform of the electric field in real space. The finite-difference time-domain method is used to compute and scan the transmission coefficient, as well as to verify the corresponding GH shift. For the full-wave simulation with a finite-sized structure, the model is enclosed by PMLs.

**Supporting Information**

Supporting Information is available from the Wiley Online Library or from the author.

**Acknowledgements**

The authors acknowledge support from National Key Research and Development Program of China (2022YFA1404800, 2023YFA1406903); National Natural Science Foundation of China (12374307, 12234009, 12427808).

Received: ((will be filled in by the editorial staff))
Revised: ((will be filled in by the editorial staff))
Published online: ((will be filled in by the editorial staff))

**Supplemental Material for "Controlling Enhancement of Transmitted Goos-Hänchen Shifts: From Symmetric to Unidirectional"**

*Zhuolin Wu, Weiming Zhen, Zhi-Cheng Ren, Xi-Lin Wang, Hui-Tian Wang and Jianping Ding*

**CONTENTS**

## S1. Explanation of Symmetry in Dislocated Layered Photonic Crystal (DLPC)

When the system satisfies inversion symmetry (*P* symmetry), the operator $\mathcal{P}$ is defined in the basis of the forward (+) and backward (–) propagation modes as $\mathcal{P} = |a_{1-}\rangle\langle a_{1+}| + |a_{1+}\rangle\langle a_{1-}|$. When generalized to an *N*-layer system, each layer having two modes for forward and backward propagation, the corresponding general formula is

$$\mathcal{P}^{(N)} = \sum_{j=1}^{N} |a_{N+1-j,-}\rangle\langle a_{j,+}| + |a_{N+1-j,+}\rangle\langle a_{j,-}|, \tag{S1}$$

In this case, the symmetry ensures that the Hamiltonian of the system satisfies the commutation relation $\left[\mathbf{H}(k,\delta_w), \mathcal{P}^{(N)}\right] = 0$, indicating that this operation is independent of the degree of dislocation $\delta_w$. Physically, this reflects the simultaneous realization of two operations: inter-layer mode swapping and flipping of forward and backward modes. Consequently, the inversion operation can be expressed by matrix multiplication as

$$\mathcal{P}^{(N)} = m_x^{(N)} \cdot m_z^{(N)} = \bigoplus_{j=1}^{N} \begin{pmatrix} 0 & 1 \\ 1 & 0 \end{pmatrix} = \begin{pmatrix} 0 & 1 & 0 & 0 & \cdots & 0 & 0 \\ 1 & 0 & 0 & 0 & \cdots & 0 & 0 \\ 0 & 0 & 0 & 1 & \cdots & 0 & 0 \\ 0 & 0 & 1 & 0 & \cdots & 0 & 0 \\ \vdots & \vdots & \vdots & \vdots & \ddots & \vdots & \vdots \\ 0 & 0 & 0 & 0 & \cdots & 0 & 1 \\ 0 & 0 & 0 & 0 & \cdots & 1 & 0 \end{pmatrix}, \tag{S2}$$

where $m_x$ represents the *x*-reflection symmetry operation, which swaps the forward and backward modes within each layer. On the other hand, $m_z$, as the *z*-reflection symmetry, swaps the *j*-th layer with the (*N*+1–*j*)-th layer, corresponding to the *N* identity matrices along the anti-diagonal. Notably, unlike the preservation of *x*-reflection symmetry, the *z*-reflection symmetry imposes the condition $\mathbf{H}(k, \delta_w) = \mathbf{H}(k, -\delta_w)$, which is independent of the wavevector *k*. This leads to two sets of solutions, $\delta_w = 0$ and $\delta_w = \pm 0.5$, indicating that the band structures of both the trivial dislocation-free structure and the special half-period dislocation structure are equivalent at any position in momentum space, protected by *z*-reflection symmetry. When the multi-layer system degenerates to a single layer, the inversion symmetry degenerates to the first Pauli matrix $\boldsymbol{\sigma}_x$, consistent with *x*-reflection symmetry, while the *z*-reflection symmetry is represented by the 2nd-order identity matrix $\mathbf{I}_2$.

## S2. Constraints on the Coupling Coefficients by the Physical System

When the *z*-reflection symmetry of the structure is broken, the constraints on **D**, **K**, and **C** that held in the symmetric system are no longer valid. Therefore, it becomes necessary to rederive the relationships between the coupling coefficients in the DC-SCMT within the DLPC system. Specifically, these are given by $\mathbf{D}^\dagger \mathbf{D} = 2\gamma \mathbf{I}$, $\mathbf{K} = -\mathbf{D}\boldsymbol{\sigma}_x \boldsymbol{\alpha}_{v_g}$ and $\mathbf{C}\mathbf{D}^* = -\mathbf{D}\boldsymbol{\sigma}_x$.

### a. Proof of Constraint I.

For systems without intrinsic material absorption losses, the system always satisfies energy conservation. Consider the passive case, where $|s_+\rangle = 0$, in which only the energy decay of the mode is exhibited

$$\boldsymbol{\alpha}_{v_g} \frac{d(\mathbf{A}^\dagger \mathbf{A})}{d\mathbf{r}_\parallel} = -\mathbf{A}^\dagger (2\gamma \mathbf{I}) \mathbf{A}. \tag{S3}$$

This should be consistent with the scattered energy detected at the port

$$-\langle s_- | s_-\rangle = -\mathbf{A}^\dagger \boldsymbol{\alpha}_{v_g} \mathbf{D}^\dagger \mathbf{D} \boldsymbol{\alpha}_{v_g} \mathbf{A}. \tag{S4}$$

Therefore, the first constraint can be immediately obtained as

$$\mathbf{D}^\dagger \mathbf{D} = 2\gamma \mathbf{I}. \tag{S5}$$

This reveals the relationship between mode radiation and port coupling. In the scenario of a stacked dislocation viewed as a single layer, this corresponds to Equation (11) in the main text. Since the system satisfies *P*-symmetry, it can be simplified as $\psi_+ = \psi_-$, $d_1 = d_4$, and $d_2 = d_3$. In the case of half-period dislocations or no dislocations, the four ports degenerate to equality. When the system excites UGR, one of the pairs, either $d_1$ and $d_3$ or $d_2$ and $d_4$, will approach zero.

### b. Proof of Constraint II.

Consider the time reversal of the source-free scenario, that is, the incident state transforms as $|s_{\text{in}}\rangle = |s_{\text{out}}\rangle^*$. Symmetry ensures that the backward gain mode $\boldsymbol{\sigma}_x \mathbf{A}'$ remains a solution of the dynamical equation, which must satisfy

$$\left[i(|\mathbf{k}_\parallel| - |\mathbf{k}_0|)\boldsymbol{\sigma}_z - \gamma^* \boldsymbol{\alpha}_{v_g}\right] \boldsymbol{\sigma}_x \mathbf{A}' = \boldsymbol{\alpha}_{v_g} \boldsymbol{\sigma}_x \mathbf{K}^T |s_+\rangle, \tag{S6}$$

noting that the mode coupling ports change accordingly. Consider the wave vector $|\mathbf{k}_\parallel|\boldsymbol{\sigma}_z = |\mathbf{k}_0|\boldsymbol{\sigma}_z + i\gamma \boldsymbol{\alpha}_{v_g}$, and taking into account that $\alpha_{v_g}^+ \alpha_{v_g}^- = -1$, i.e., $-\boldsymbol{\sigma}_x \mathbf{A}' = \mathbf{A}^*$, we obtain

$$-\mathbf{D}^T \mathbf{D}^* = \boldsymbol{\alpha}_{v_g} \boldsymbol{\sigma}_x \mathbf{K}^T \mathbf{D}^* \tag{S7}$$

where the first constraint has been applied. Subsequently, since the radiation of modes to different ports is correlated, it is necessary to decouple the port dissipation using singular value decomposition (SVD).[1] At this point, the coupling coefficients can be rewritten as

$$\mathbf{D} = \mathbf{U} \begin{bmatrix} \mathbf{D} & 0 \\ 0 & 0 \end{bmatrix} \mathbf{V}^\dagger, \mathbf{K} = \mathbf{U} \begin{bmatrix} \mathbf{K} & 0 \\ 0 & 0 \end{bmatrix} \mathbf{V}^\dagger \tag{S8}$$

Substituting into Equation (S7), excluding trivial solution results in $\mathbf{K} = -\mathbf{D}\boldsymbol{\sigma}_x \boldsymbol{\alpha}_{v_g}$. Thus,

$$\mathbf{K} = \mathbf{U}\begin{bmatrix} \mathbf{K} & 0 \\ 0 & 0 \end{bmatrix}\mathbf{V}^\dagger = \mathbf{U}\begin{bmatrix} -\mathbf{D}\boldsymbol{\sigma}_x \boldsymbol{\alpha}_{v_g} & 0 \\ 0 & 0 \end{bmatrix}\mathbf{V}^\dagger = -\mathbf{D}\boldsymbol{\sigma}_x \boldsymbol{\alpha}_{v_g} \tag{S9}$$

This is the second constraint. It highlights the interaction between the modes and the energy coupling of input and output at the ports.

### c. Proof of Constraint III.

Time-reversal symmetry, in addition to its effects on the dynamical equations, also provides insights based on the scattering equations. Considering that radiation is not allowed in this case, we can directly write

$$\mathbf{0} = \mathbf{C}|s_-\rangle^* + \mathbf{D}\boldsymbol{\sigma}_x \mathbf{A}^* = \mathbf{C}\mathbf{D}^*\boldsymbol{\alpha}_{v_g}\mathbf{A}^* + \mathbf{D}\boldsymbol{\sigma}_x \boldsymbol{\alpha}_{v_g}\mathbf{A}^*. \tag{S10}$$

Therefore, the coupling coefficients must satisfy

$$\mathbf{C}\mathbf{D}^* = -\mathbf{D}\boldsymbol{\sigma}_x, \tag{S11}$$

which establishes a direct relationship between scattering and mode radiation.

### S3. Derivation of the Full Transmission Condition in DLPC Based on DC-SCMT

To specifically demonstrate the advantages of our transmitted shift scheme in terms of transmission efficiency, within our theoretical framework, we consider the example of a single-layer four-port dual-resonance model. Following the treatment similar to the reference[2], we start with the transmission and reflection coefficients under incidence from port 1

$$t_{41} = it + \frac{d_4 d_1 \exp\left[i(\psi_- + \psi_+)/2\right]}{i(|k_\parallel| - |k_0|) + \alpha_{v_g}^+ \gamma}, \quad r_{21} = r + \frac{d_1 d_2 \exp\left[i(\psi_- - \psi_+)/2\right]}{i(|k_\parallel| - |k_0|) + \alpha_{v_g}^+ \gamma} \tag{S12}$$

We redefine the radiation coupling coefficients as $d_j' = d_j \exp(i\psi_j/2) = \sqrt{2\gamma_j'}\exp(i\xi_j')$, where $j = $ 1-4, and denote $\alpha = \xi_4' + \xi_1', \beta = \xi_1' + \xi_2'$. Expanding the third constraint reveals two independent equations

$$\begin{aligned}-\sqrt{2\gamma_3'}\exp(i\xi_3' - i\xi_1') &= r\sqrt{2\gamma_4'}\exp(-i\alpha) + it\sqrt{2\gamma_2'}\exp(-i\beta), \\ -\sqrt{2\gamma_1'} &= it\sqrt{2\gamma_4'}\exp(-i\alpha) + r\sqrt{2\gamma_2'}\exp(-i\beta).\end{aligned} \tag{S13}$$

The goal now is to solve for the exponential of $\alpha$ to express the transmission coefficient. Therefore, the terms related to $\beta$ are expressed in terms of $\alpha$, i.e., $r\sqrt{2\gamma_2'}\exp(-i\beta) = -\sqrt{2\gamma_1'} - it\sqrt{2\gamma_4'}\exp(-i\alpha)$, and by taking the squared norm, we obtain

$$2r^2\gamma_2' = 2\gamma_1' + 2t^2\gamma_4' + 4t\sqrt{\gamma_4'\gamma_1'}\sin\alpha. \tag{S14}$$

From the expression above, we can derive the trigonometric functions of $\alpha$

$$\sin\alpha = \frac{\delta - t^2\gamma}{2t\sqrt{\gamma'_4\gamma'_1}},$$
$$\cos\alpha = \pm\frac{\sqrt{4\gamma'_4\gamma'_1 - \delta^2/t^2 - t^2\gamma^2 + 2\delta\gamma}}{2\sqrt{\gamma'_4\gamma'_1}},$$
(S15)

where, considering that $\gamma = \gamma'_2 + \gamma'_4$, $\alpha^+_{v_g} = 1$, and we define $\delta = \gamma'_2 - \gamma'_1$. Thus, the transmission is rewritten as

$$T_{41} = |t_{41}|^2 = \left|it + \frac{2\sqrt{\gamma'_4\gamma'_1}\exp(i\alpha)}{i(|k_\parallel| - k_0) + \alpha^+_{v_g}\gamma}\right|^2$$
$$= \frac{t^2(|k_\parallel| - k_0)^2 + t^2\gamma^2 + 4\gamma'_4\gamma'_1 - 4t\sqrt{\gamma'_4\gamma'_1}(|k_\parallel| - k_0)\cos\alpha + 4t\gamma\sqrt{\gamma'_4\gamma'_1}\sin\alpha}{(|k_\parallel| - k_0)^2 + \gamma^2},$$
(S16)

where Euler's formula has been applied. When total transmission occurs, i.e., $T_{41} = 1$, this is equivalent to

$$(1-t^2)(|k_\parallel| - k_0)^2 + (1-t^2)\gamma^2 - 4\gamma'_4\gamma'_1 + 4t\sqrt{\gamma'_4\gamma'_1}(|k_\parallel| - k_0)\cos\alpha - 4t\gamma\sqrt{\gamma'_4\gamma'_1}\sin\alpha = 0 \quad (S17)$$

The discriminant of the above equation can be expressed as

$$\Delta = 16t^2\gamma'_4\gamma'_1\cos^2\alpha - 4(1-t^2)(1-t^2)\gamma^2 + 16(1-t^2)\gamma'_4\gamma'_1 + 16(1-t^2)t\gamma\sqrt{\gamma'_4\gamma'_1}\sin\alpha$$
$$= -4(\gamma'_1 - \gamma'_4)^2,$$
(S18)

where Equation (S15) has been substituted. Therefore, the equation has a solution only when $\gamma'_1 = \gamma'_4$, which indicates that the $P$-symmetry characteristic of the port makes total transmission possible.

**S4. Band Structure and Polarization Field of Mode D during the Evolution of Layer Spacing**

Near Mode D, the bands are observed to approach each other as the interlayer spacing increases, as shown in **Figure S1**(a). The dashed lines clearly indicate the progressively narrowing gap between the $TE_1$ band and the band above it, which suggests increasing crosstalk between the two modes in the scattering response. As a result, even the UGR on the $TE_1$, which suppress radiation, cannot achieve a smooth and linear phase transition. This is the critical limitation mentioned in the main text that constrains the displacement asymmetry factor. During the evolution shown in Figure S1(a), two downward radiation-suppressed UGRs are sequentially encountered, corresponding to Modes D and E, as marked in Figure 4(b) in the main text. The polarization field distributions of the two UGRs, shown in Figures S1(b) and (c), both exhibit the topological characteristics of a V-point on the down side.

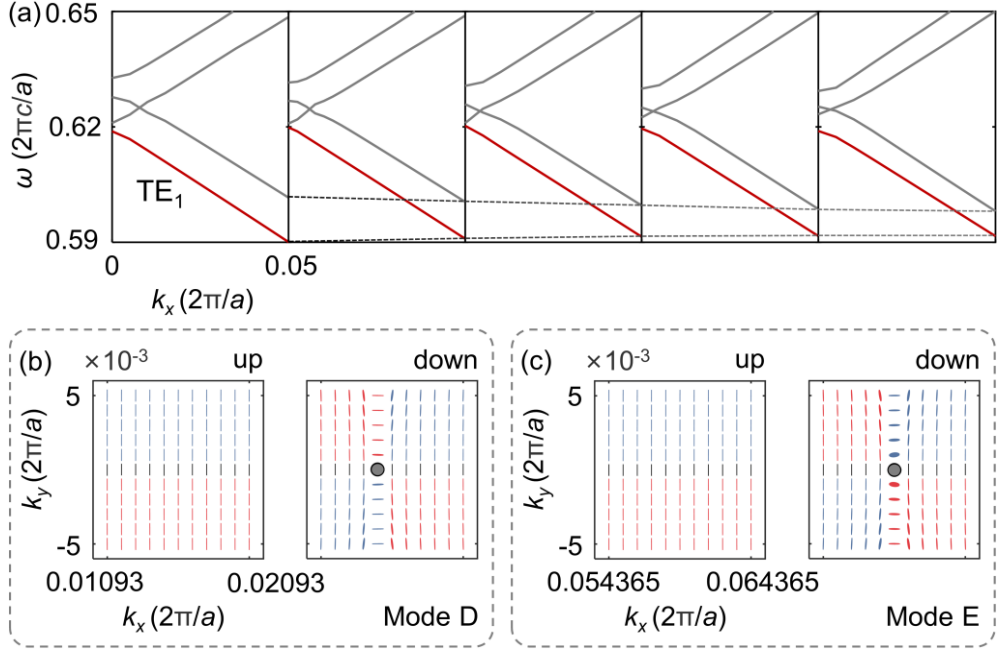

**Figure S1.** (a) For the DLPC with interlayer spacing, the evolution of the band structure near Mode D as the interlayer spacing increases. The TE$_1$ band of interest is highlighted in red. (b) and (c) show the polarization field distributions for Modes D and E in the main text on the upper and lower sides, respectively, where red represents LCP, blue represents RCP, and black indicates linear polarization.

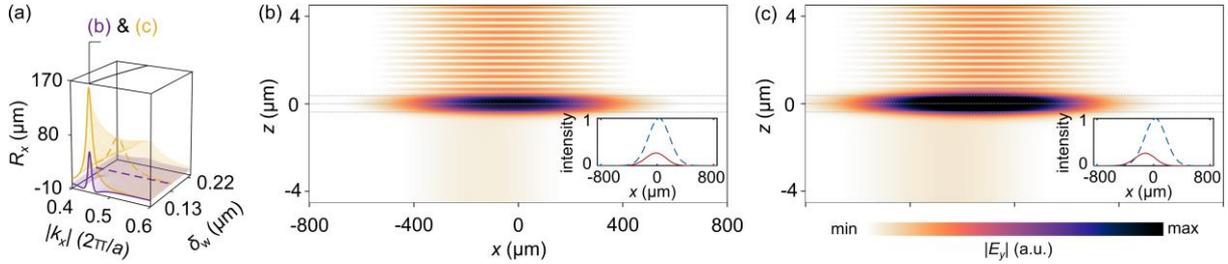

**Figure S2.** (a) Displacement distribution obtained from oblique incidence on the left (purple) and right (yellow) sides of the parameter space for the stacked DLPC, with the dashed line indicating the corresponding $\delta_w$ for the unidirectional enhanced shift studied in the main text. The upper plane extracts the evolution trajectory of the maximum displacement. Corresponding to the solid line in Figure. S2(a), (b) and (c) illustrate the field distribution at the cross-section where the GH shift is maximum for $\delta_w = 0.0357a$, for left- and rightward oblique incidence, respectively. The corresponding insets show the intensity slices at the incident port (blue dashed line) and the outgoing port (red solid line). To thoroughly excite the GH shift, a Gaussian beam with a waist of $w_0 = 300$ μm is considered for the discussion.

## S5. Verification of the Goos-Hänchen (GH) Shift during Dislocation Evolution

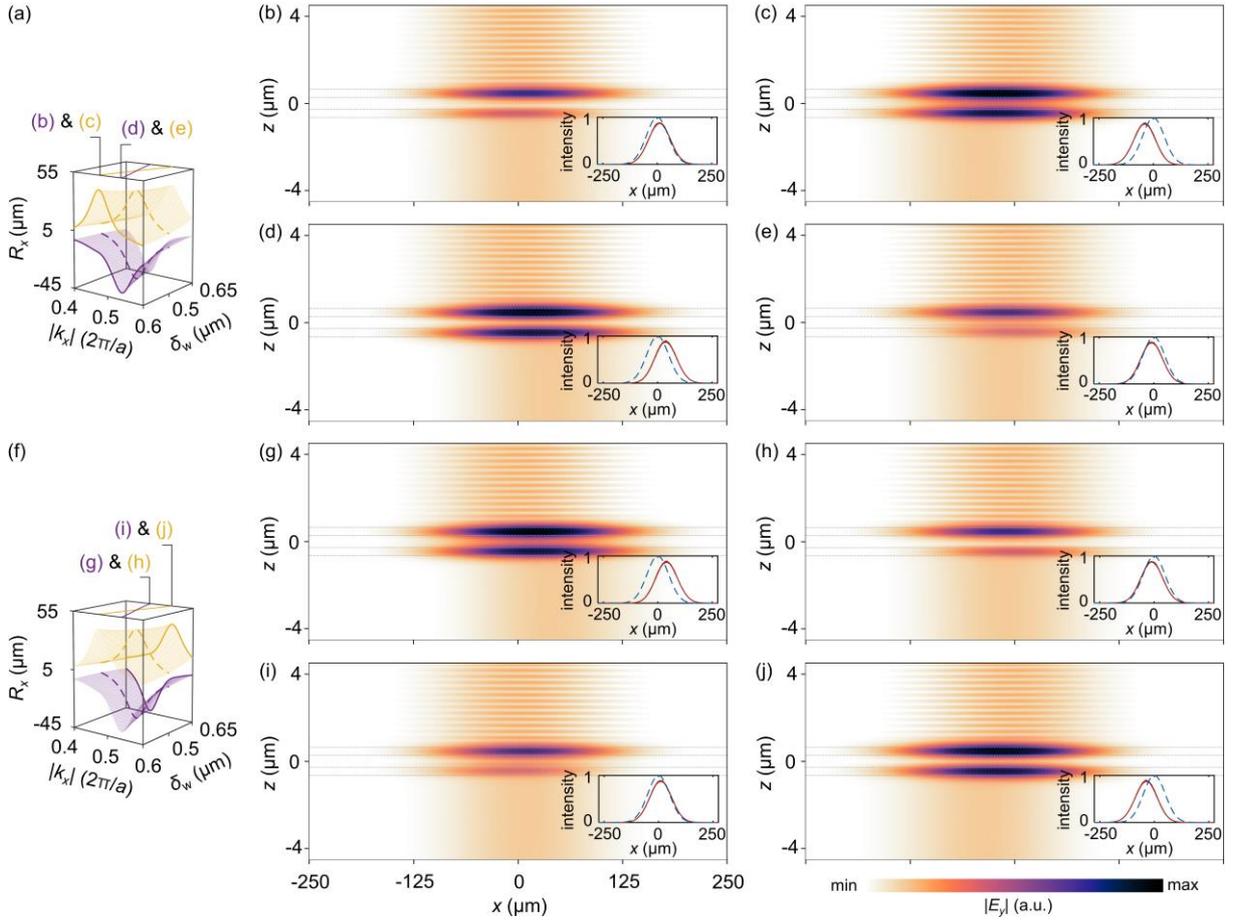

**Figure S3.** For the half-period DLPC with fixed interlayer spacing, (a) and (f) show the displacement distribution from oblique incidence on the left (purple) and right (yellow) sides of the parameter space. The dashed line indicates the corresponding $\delta_w$ for the symmetric shift studied in the main text. The upper plane extracts the evolution trajectory of the maximum displacement. For the solid line in Figure. S3(a), when $\delta_w = 0.35a$, (b) and (c) represent the field distributions of the GH shift cross-section from the perspective of the wavenumber corresponding to the maximum forward shift, obtained by oblique incidence to the left and right, respectively. Meanwhile, (d) and (e) correspond to the response of the wavenumber matching the maximum backward shift. In Fig. S3(f), for the solid line and with $\delta_w = 0.65a$, (b) and (c) show the field distributions of the GH beam shift cross-section, corresponding to the wavenumber associated with the maximum backward shift, for left- and rightward oblique incidence, respectively. In contrast, (i) and (j) present the response for the wavenumber corresponding to the maximum forward shift. The corresponding insets depict the intensity slices at the incident port (blue dashed line) and the outgoing port (red solid line). To thoroughly excite the GH shift, a Gaussian beam with a waist of $w_0 = 100$ μm is considered for the discussion.

**Figure S2**(a) provides supplementary validation for Figure 2 in the main text, specifically showing the parameter space distribution near the position of the unidirectional enhanced shift. At this point, the maximum displacement evolution trajectories for both forward and backward shifts coincide. Interestingly, contrary to intuition, the displacement directions obtained from oblique incidence at different directions are the same. This is due to the weak coupling between the mode and the background spectrum, which is demonstrated through the field distributions in Figures S2(b) and (c). Combining with Figure 2(b) in the main text, the lower transmission corresponds to the reduced field strength at the outgoing port, as shown in the insets of Figures S2(b) and (c). Since the incident beam waist radius is sufficiently large, the resulting GH shift closely matches the approximate result given in equation (2) of the main text, with displacement values of 47.601 μm and 151.656 μm for Figures S2(b) and (c), respectively.

**Figures S3**(a) and (f) provide supplementary validation for Figure 6 in the main text, specifically illustrating the parameter space distribution near the position of the symmetric shift enhancement. At this point, the maximum GH shift evolution trajectories for the forward and backward no longer overlap. For smaller values of $\delta_w$, the displacements corresponding to the wavevectors matching the maximum forward shift are 10.564 μm (left, Figure S3(b)) and 41.5992 μm (right, Figure S3(c)), while the displacements for the wavevectors matching the maximum backward shift are 38.311 μm and 9.97025 μm, as shown in Figures S3(d) and (e), respectively. Similarly, for larger $\delta_w$, the magnitudes of the two opposite-direction GH shift for different wavevectors exhibit a reversal compared to the previous results. This can be attributed to the equivalence of the scattering responses between the half-period dislocated and non-dislocated structures, where the responses before and after the reversal are symmetric. Additionally, since the waist radius of the incident beam is sufficiently large, the GH shifts obtained here closely approximate the result from Equation (2) in the main text.

**S6. Environment Refractive Index Sensor Based on Giant Transmitted GH Shifts**

As an application, the unidirectional enhanced transmitted GH shift is highly suitable for environmental refractive index sensing. In contrast to the typical scenarios of GH shifts in reflection, the realization of enhanced transmitted shifts has always been challenging. However, our unidirectional GH shift enhancement scheme, based on DLPC, overcomes the issue of reduced transmission efficiency during phase transitions. This provides an effective and feasible solution for integrated, high-efficient, and high-sensitive sensing in transmission configurations. **Figure S4**(a) illustrates that as the environmental refractive index changes, the peak position of the GH shift moves within the angular spectrum, resulting in a variation in the displacement at the original resonance position. This variation, $\Delta n$, serves as the basis for high-sensitivity sensing. In Figure S4(b), by fixing the wave vector corresponding to the peak of GH shift at $n_b$ = 1.5, the GH shift associated with this wave vector gradually decreases as the refractive index increases. Consequently, the sensitivity, defined as $S = |d\delta_x/dn|$, first increases and then decreases, with a theoretical maximum of approximately $6.570\times10^6$ nm/RIU, while remaining above $10^6$ nm/RIU over a wide range.

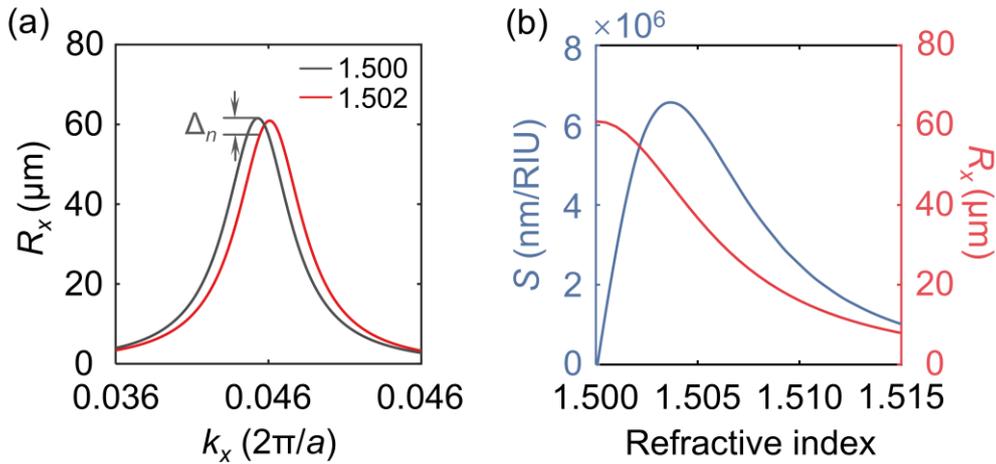

**Figure. S4.** (a) The response of unidirectional shift enhancement for environmental refractive indices of 1.500 and 1.502, respectively. (b) The variation of the maximum beam shift and its refractive index-dependent sensitivity as the environmental refractive index changes. The structure and light sssource are set according to the parameters corresponding to Mode C in the main text.